\documentstyle[emulateapj]{article}

\lefthead{Nagao et al.}
\righthead{Mrk 573}

\begin{document}

\submitted{The Astronomical Journal, in Press} 
\title{Detection of the Polarized Broad Emission Line 
       in the Seyfert 2 Galaxy Mrk 573\footnote{
         Based on data collected at Subaru Telescope, which is operated
         by the National Astronomical Observatory of Japan.}
       }

\author{Tohru NAGAO$^2$, Koji S. KAWABATA$^3$, Takashi MURAYAMA$^2$, 
        Youichi OHYAMA$^4$, Yoshiaki TANIGUCHI$^2$, \\
        Ryoko SUMIYA$^2$, and Shunji S. SASAKI$^2$}
\affil{$^2$ Astronomical Institute, Graduate School of Science, 
       Tohoku University, Aramaki, Aoba, Sendai 980-8578, Japan}
\affil{$^3$ Department of Physical Science, 
       Graduate School of Science, Hiroshima University, 
       1-3-1 Kagamiyama, Higashi-Hiroshima, \\
       Hiroshima 739-8526, Japan}
\affil{$^4$ Subaru Telescope, National Astronomical Observatory of
       Japan, 650 North A`ohoku Place, University Park, Hilo, HI 96720}


\begin{abstract}

We report the discovery of the scattered emission from a hidden broad-line
region (BLR) in a Seyfert 2 galaxy, Mrk 573, based on our recent 
spectropolarimetric observation performed at the Subaru Telescope.
This object has been regarded as a type 2 AGN without a hidden BLR
by the previous observations. However, our high quality
spectrum of the polarized flux of Mrk 573 shows prominent broad 
($\sim$3000 km s$^{-1}$) H$\alpha$ emission, broad weak H$\beta$ emission, 
and subtle Fe {\sc ii} multiplet emission.
Our new detection of these indications for the presence of the hidden BLR 
in the nucleus of Mrk 573 is thought to be owing to the high
signal-to-noise ratio of our data, but the possibility of 
a time variation of the scattered BLR emission is also mentioned.
Some diagnostic quantities such as the $IRAS$ color, the radio
power, and the line ratio of the emission from the narrow-line region
of Mrk 573 are consistent with the distributions of such quantities of 
type 2 AGNs with a hidden BLR. Mrk 573 is thought to be an object
whose level of the AGN activity is the weakest among the type 2 AGNs with
a hidden BLR. In terms of the systematic differences
between the type 2 AGNs with and without a hidden BLR,
we briefly comment on an interesting Seyfert 2 galaxy,
Mrk 266SW, which may possess a hidden BLR but has been treated as a
type 2 AGNs without a hidden BLR.

\end{abstract}



\section{INTRODUCTION}

\begin{deluxetable}{lcccc}
\tablenum{1}
\tablecaption{Polarization Properties of Stars near the Line-of-Sight
              of Mrk 573\tablenotemark{a}}
\tablewidth{0pt}
\tablehead{
\colhead{Star Name} &
\colhead{Separation} &
\colhead{Distance} &
\colhead{Pol. Degree} &
\colhead{Pol. Angle} \\
\colhead{} &
\colhead{(arcmin)} &
\colhead{(parsec)} & 
\colhead{(\%)} &
\colhead{(deg)} 
}
\startdata  
HD  9740 &  103  &  436  & 0.33 & 122.5 \nl
HD 10441 &  155  &  316  & 0.30 & 128.4 \nl
adopted  &\nodata&\nodata& 0.32 & 125
\enddata 
\tablenotetext{a}{Data are taken from Heiles (2000).}
\end{deluxetable}

After the discovery of scattered broad permitted lines in the
polarized spectra of a Seyfert 2 galaxy (hereafter Sy2) NGC 1068 
(Antonucci \& Miller 1985), many attempts have been made to detect
the polarized broad lines in type-2 active galactic nuclei (AGNs) up 
to now (e.g., Miller \& Goodrich 1990; Tran, Miller, \& Kay 1992;
Young et al. 1993; Tran, Cohen, \& Goodrich 1995; Tran 1995; 
Young et al. 1996; Kay \& Moran 1998; 
Barth, Filippenko, \& Moran 1999a, 1999b; 
Tran, Cohen, \& Villar-Martin 2000; Kishimoto et al. 2001; Tran 2001;
Lumsden et al. 2001).
This is because the presence of scattered broad permitted lines is
promising evidence for the AGN unified model,
in which both the type-1 and type-2 AGNs possess a broad-line region
(BLR) in their nucleus (see Antonucci 1993 for a review).

Although the scattered BLR emission has been found in many type-2
AGNs, it is now recognized that Sy2s do not always exhibit
broad lines in the polarized spectra (see, e.g., Tran 2001).
Does this suggest that not all Sy2s possess a BLR in their nucleus?
This is very important issue because the presence of the two distinct 
populations of Sy2s, i.e., Sy2s with and without a BLR, is 
inconsistent to the simple AGN unified model.
Heisler, Lumsden, \& Bailey (1997) proposed that this dichotomy 
can be understood in the framework of the unified model
if the scattering region resides very close to the nucleus and 
its visibility depends on the viewing angle (see also, e.g.,
Taniguchi \& Anabuki 1999).
On the other hand, Tran (2001) reported that the amount of the 
obscuration toward the central engine is indistinguishable between 
the Sy2s with and without polarized BLR emission (see also
Alexander 2001).
This suggests the presence of AGNs without BLRs (see, e.g.,
Heckman et al. 1995; Dultzin-Hacyan et al. 1999; 
Gu, Maiolino, \& Dultzin-Hacyan 2001), 
which contradicts the current simple unified model.
Investigating this issue further is crucially important not only to
examine the AGN unified model but also to understand the nature of 
AGN phenomena themselves.

In order to discuss this issue, we should recognize correctly
which Sy2 possesses a hidden BLR and which Sy2 does not.
Here we report a clear detection of the hidden BLR in a Sy2,
Mrk 573, which has been regarded as a Sy2 {\it without} a hidden BLR
(Tran 2001, 2003; see also Kay 1994). 
Its heliocentric radial velocity is
5156 $\pm$ 90 km s$^{-1}$ (Whittle et al. 1988), giving a projected
linear scale of 0.33 $h_{75}^{-1}$ kpc for 1 arcsec.

\section{OBSERVATION AND DATA REDUCTION}

The spectropolarimetric observation for Mrk 573 was carried out by using 
FOCAS, Faint Object Camera And Spectrograph (Kashikawa et al. 2002) 
on the 8.2m Subaru Telescope (Kaifu 1998) at Mauna Kea,
on 2003 October 5--6 (UT). The FOCAS detector is a mosaic of two 
4k $\times$ 2k MIT CCDs with 15$\mu$m pixels.
All the observations were carried out through a polarimetric unit that 
consists of a rotating superachromatic half-wave plate and a quartz 
Wollaston prism. A 0.4$^{\prime \prime}$ width slit, 
a 300 lines mm$^{-1}$ grisms (300B), and an order-sorting filter of 
Y47 were used.
This setting results in a wavelength resolution of $R \sim 1000$.
We adopted a 3-pixel binning for the spatial direction on the chips,
which results in the spatial sampling rate of 0.31 arcsec for a
binned pixel.
All of the data are obtained at four wave-plate position angles,
0.0$^{\circ}$, 45.0$^{\circ}$, 22.5$^{\circ}$, and 67.5$^{\circ}$.
The integration time of each exposure for the observation of Mrk 573
is 240 or 480 seconds, and the total on-source integration time is
208 minutes. The position angle of the slit was set to
be 0$^{\circ}$.
We also obtained spectra of unpolarized standard stars
(BD+28$^{\circ}$ 4211 and G191B2B) and a strongly polarized star (HD 204827).
Spectra of a halogen lamp and a thorium-argon lamp were also obtained
for the flat fielding and the wavelength calibration, respectively.

The data were reduced by the standard manner, by using IRAF\footnote{
IRAF (Image Reduction and Analysis Facility) is distributed by the
National Optical Astronomy Observatory, which is operated by the
Association of Universities for Research in Astronomy Inc., under 
corporative agreement with the National Science Foundation.}.
We extracted the spectra of Mrk 573 and the standard stars
by adopting the aperture size of 3.1
arcsec (i.e., 10 binned pixels).
The corresponding linear aperture size in the frame of Mrk 573
is 1.02 $h_{75}^{-1}$ kpc $\times$ 0.13 $h_{75}^{-1}$ kpc.
The instrumental polarization was corrected by using the data of
the unpolarized standard stars. The instrumental depolarization was
not corrected because it has been confirmed experimentally that the
amount of the instrumental depolarization of the FOCAS is less than 
a few percent. The flux calibration was performed by using the data
of BD+28$^{\circ}$ 4211 and G191B2B (Oke 1990).
The polarization angle was calibrated by using the data of the strongly
polarized star.

The Galactic interstellar polarization toward the direction of
Mrk 573 is estimated to be $P=0.32$\% and $\theta=125$ deg at $B$ band,
based on the polarimetric properties of the two stars near the line of
sight toward Mrk 573; i.e., HD 9740 and HD 10441 (Table 1). 
Accordingly, the obtained spectrum was corrected for the Galactic 
interstellar polarization by adopting a
Serkowski law (Serkowski, Mathewson, \& Ford 1975) with $P_{\rm max}$ 
occurring at $\lambda_{\rm max}=5500{\rm \AA}$.
Note that the interstellar polarization might be over-estimated because
the Galactic reddening toward the direction of Mrk 573 estimated by
the dust emissivity map is not so large;
i.e., $E_{B-V} = 0.023$ mag (Schlegel, Finkbeiner, \& Douglas 1998).
Since the Galactic interstellar polarization
is $\sim$3 $E_{B-V}$ \% for typical conditions and $\sim$9 $E_{B-V}$ \%
at most (see Serkowski et al. 1975), the Galactic interstellar polarization 
for Mrk 573 is estimated to be $\lesssim$0.21 \% which contradicts our
adopted value, $P=0.32$\%.
However, this discrepancy may be due to the underestimation of the 
Galactic reddening. The spectroscopically determined Galactic reddening
for HD 9740 and HD 10441 are both $E_{B-V} = 0.10$ mag (Heiles 2000),
about three times larger than that estimated from the dust emissivity map
($E_{B-V} = 0.03$ mag; Schlegel et al. 1998).
Therefore, as for the sky region around Mrk 573,
there may be a tendency that the Galactic reddening is underestimated
when using the dust emissivity map.
Note that, in any cases, this uncertainty does not affect the following 
discussion about the detection of the broad
component of Balmer lines, which is the main concern of this paper.

\section{RESULTS}

\begin{figure*}
\epsscale{1.4}
\plotone{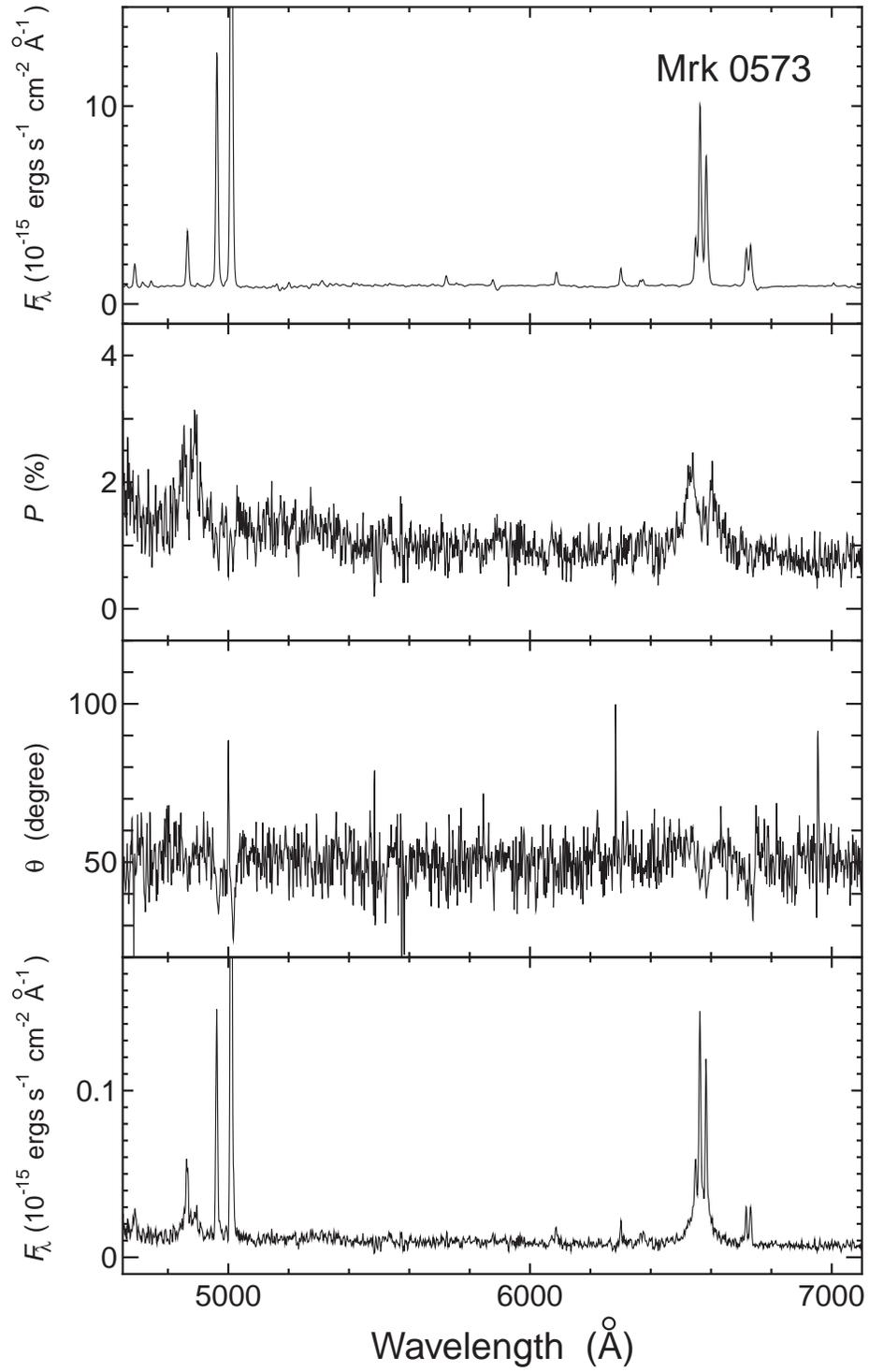}
\caption{
The obtained data of Mrk 573 are plotted as a function of wavelength.
The data are corrected for the interstellar polarization but
not corrected for reddening, redshift, and starlight of the host galaxy;
(a) the total flux, $I$,
(b) the polarization degree, $P$,
(c) the position angle of polarization, $\theta$, and
(d) the polarized flux (i.e., $I \times P$).
\label{fig1}}
\end{figure*}

\begin{figure*}
\epsscale{1.19}
\plotone{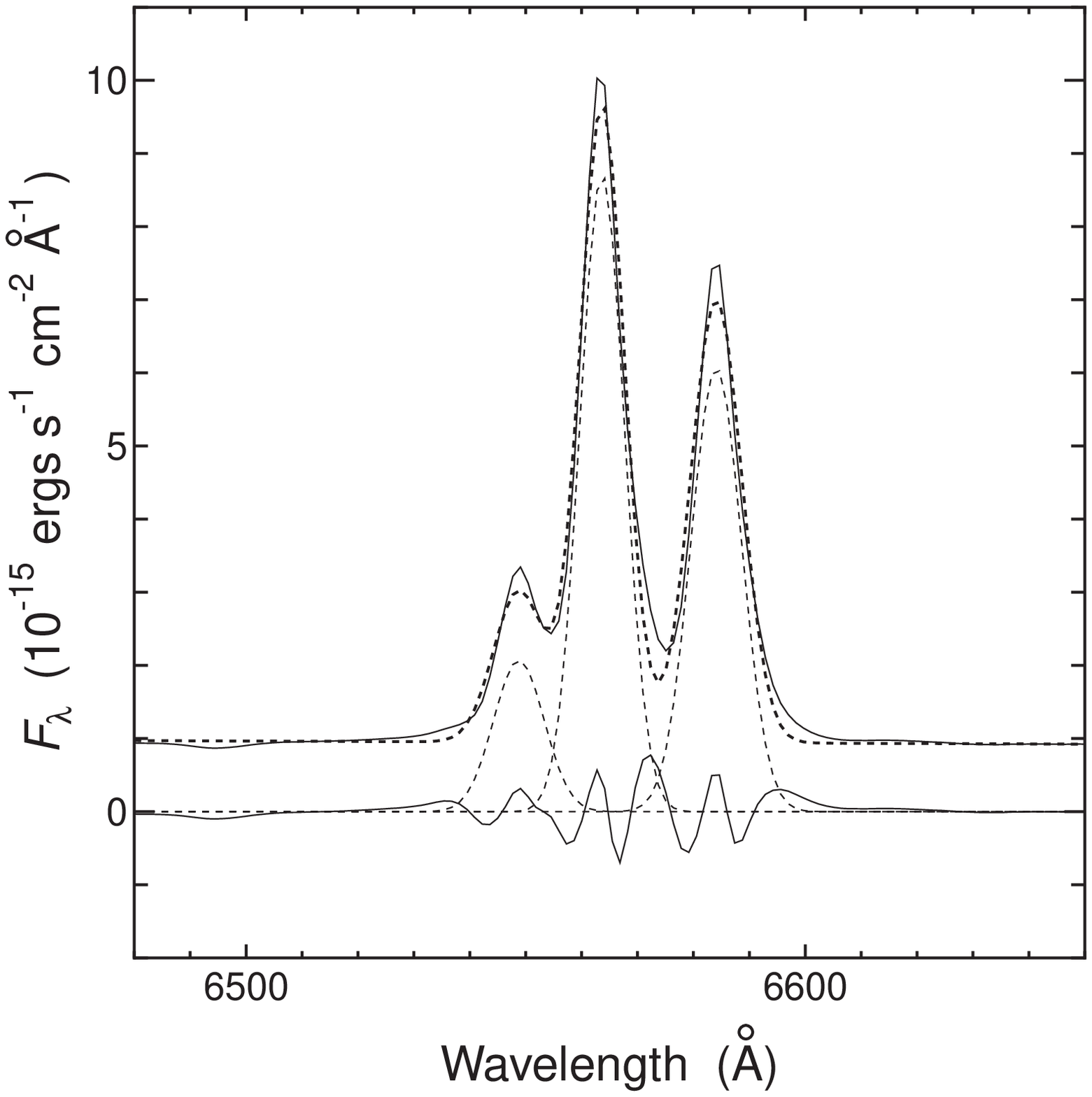}
\caption{
The obtained spectrum (solid line) and the results of the spectral
fitting (dotted lines) of the total flux, around the H$\alpha$ emission.
The dotted thin lines are three Gaussian components in the fitting model,
and the dotted thick line is the sum of the three Gaussian and the 
power-law continuum of the best-fit model. The residual spectrum is also
shown.
\label{fig2}}
\end{figure*}

The total flux ($I$), the polarization degree ($P$), the position angle 
of polarization ($\theta$), and the polarized flux ($I \times P$)
of Mrk 573 are shown as a function of wavelength in Figure 1. 
The spectra displayed in this figure are
uncorrected for reddening and starlight of the host galaxy, but
corrected for the redshift and the Galactic interstellar polarization 
as described in \S2.
Since the optical spectrum of Mrk 573 is significantly contaminated
with the starlight of the host galaxy ($\sim$80\% of the continuum;
Kay 1994), the presented spectrum of the polarization degree (Figure 1b)
is heavily diluted. However, we can investigate the spectropolarimetric
properties of AGNs through spectra of a polarized flux, because the
starlight emission of host galaxies can be regarded as unpolarized light
and thus spectra of the polarized flux are not influenced by host galaxies
(see, e.g., Antonucci \& Miller 1985; Miller \& Goodrich 1990).

The polarization degree of the continuum emission of Mrk 573 is 
1.0 $\pm$ 0.2 \%, which is measured in the wavelength range of 
5400${\rm \AA}$ -- 6000${\rm \AA}$ in the rest frame of Mrk 573,
where no strong emission-line features are present. 
The polarization angle is roughly constant at all the wavelength 
coverage, i.e., $\theta = 50.5 \pm 7.1$ deg (measured at
5000${\rm \AA}$ -- 7000${\rm \AA}$), as shown in Figure 1c.
These results are consistent with the previous observations;
e.g., Kay (1994) reported $P = 1.3$\% and $\theta =$ 48 deg
based on their spectropolarimetry.
The observed polarization angle is roughly perpendicular to the 
extension of the nuclear radio jet ($\sim 125$ deg; e.g.,
Ulvestad \& Wilson 1984). This is the same trend as that seen in many 
Sy2s (e.g., Antonucci 1983; Brindle et al. 1990).
As shown in Figure 1 clearly, narrow emission lines such as 
[O {\sc iii}]$\lambda$5007 are polarized and thus we can see the 
emission-line features in the spectrum of the polarized flux.
We do not discuss the properties of these polarized narrow emission lines
because they are investigated in the subsequent paper.

\begin{figure*}
\epsscale{1.19}
\plotone{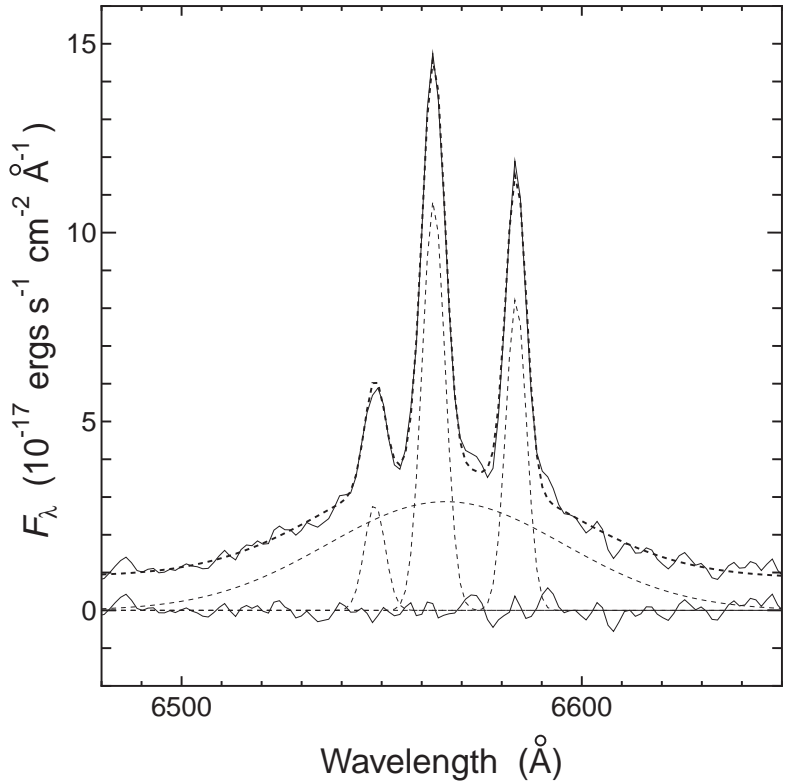}
\caption{
Same as Figure 2 but for the polarized flux. Additional broad Gaussian 
component is included in the model fitting.
\label{fig3}}
\end{figure*}

Here we focus on the polarization properties of the hidden BLR in Mrk 573,
the main concern of this paper. As shown in Figure 1, a prominent broad
component of the H$\alpha$ emission is clearly detected in the spectrum of
the polarized flux although there is no corresponding broad H$\alpha$
feature in the spectrum of the total flux. 
To present the properties of this hidden BLR component more clearly, 
we show the enlarged spectrum of the total flux and of the polarized flux
around the H$\alpha$ emission in Figures 2 and 3, respectively.
In addition to the observed spectra, the results of the spectral fitting 
by multi Gaussian components are also plotted in these figures.
For the spectral fitting, we use the task SPECFIT in the IRAF
developed by Kriss (1994).
As for the total flux, the observed spectrum can be modeled by
four components; a power-law continuum and three Gaussian components
which represent the H$\alpha$ emission and the 
[N {\sc ii}]$\lambda \lambda$6548,6583 doublet (Figure 2).
Here the wavelength separation and the flux ratio of the [N {\sc ii}] 
doublet are fixed to be the theoretical values.
See Table 2 in which the best fit parameters are given.
Since the velocity widths of the three emission-line components are 
$\sim$300 -- 400 km s$^{-1}$ in FWHM,
these emission lines are thought to arise at the narrow-line region
(NLR) in Mrk 573. There is no apparent BLR emission feature in the
residual spectrum shown in Figure 2. Instead, there is unnegligible 
residual flux, which may be because the fitting by a single Gaussian
component for each line is too simple to model the NLR emission,
and/or because the spectral features of starlight such as a Balmer 
absorption affect the fitting process. 
As for the polarized flux, on the other hand, the observed spectrum
can be well reproduced by five components, i.e., a power-law component,
three Gaussian components for the NLR emission, and a broad Gaussian
component (Figure 3; Table 2). 
The velocity width of this broad component is 
$\sim$3000 km s$^{-1}$ in FWHM, which is typical for a width of the
broad H$\alpha$ component seen in spectra of Seyfert 1 galaxies (Sy1s).
We thus conclude that the hidden BLR is surely harbored in Mrk 573,
being different from the previous reports (Tran 2001; see also Kay 1994).
Note that Tran (1995) reported that the scattered BLR emission tends to 
shift blueward relative to the [O {\sc iii}] emission (i.e., systemic 
receding velocity of the object) in general. 
As for Mrk 573, however, there is no significant difference in the 
recession velocity between the scattered BLR emission and the NLR emission.

It should be also remarked that there are some other spectral features
suggesting the presence of the hidden BLR in the nucleus of Mrk 573,
than the broad H$\alpha$ component. 
One is the polarized broad H$\beta$ component, which is clearly seen in
the spectrum of the polarized flux. The presence of this scattered
broad H$\beta$ is prominently exhibited in the spectrum of the polarization
degree (Figure 1b). However we do not make attempts to fit the spectrum
for H$\beta$ since the data quality is not sufficiently high for the 
spectral modeling.
Another interesting feature is the weak hump seen from 
$\sim$5100${\rm \AA}$ to $\sim$5400${\rm \AA}$ in the spectra of 
the polarization degree and the polarized flux (Figures 1b, 1d).
This feature is thought to be the Fe {\sc ii} multiplet emission arising
at the BLR, which is not seen in the spectrum of the total flux but
is seen only in the scattered light.
In the spectrum of the polarization degree, there might be a possible 
broad component at $\sim$5900${\rm \AA}$ although the counterpart is not
clearly seen in the spectrum of the polarized flux. This may be the scattered
broad He {\sc i} $\lambda$5876 component, but the detection 
is very marginal.
To see these possible hidden-BLR features visually, we show the enlarged
spectra of the total flux, the polarization degree, and the polarized flux,
in Figure 4.

\begin{figure*}
\epsscale{1.02}
\plotone{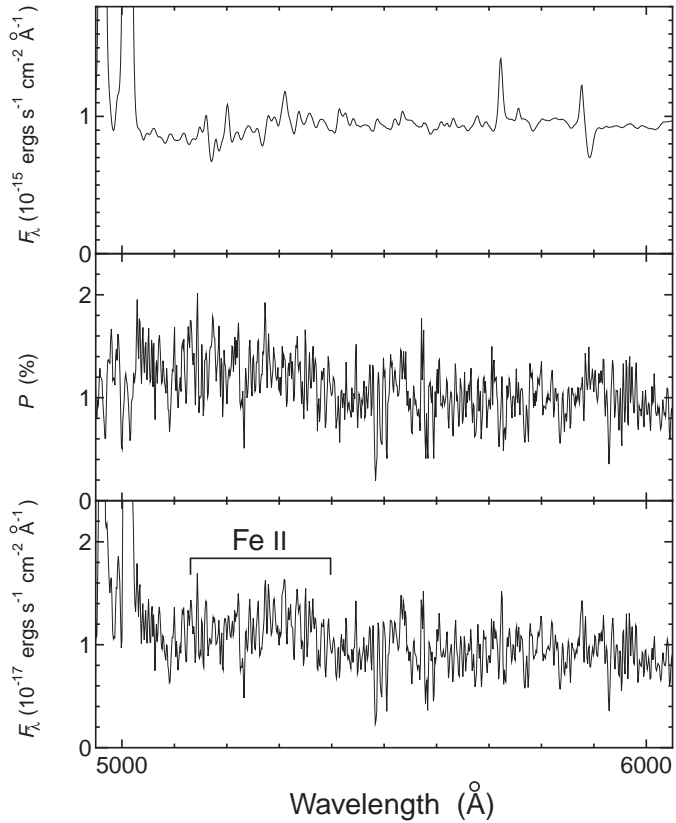}
\caption{
The enlarged spectra of the total flux, the polarization degree, and 
the polarized flux around the Fe {\sc ii} multiplet emission.
\label{fig4}}
\end{figure*}

\section{DISCUSSION}

To examine the AGN unified model is one of the most interesting issues
toward understanding the nature of the AGN phenomena.
Thus various comparative studies between Sy2s with and without the 
polarized BLR emission have been performed up to now (e.g., 
Heisler et al. 1997; Awaki et al. 2000; Gu et al. 2001; Thean et al. 2001;
Lumsden et al. 2001; Tran 2001, 2003; Gu \& Huang 2002).
In such studies, Mrk 573 has been regarded as a Sy2 without a hidden BLR.
However it is now evident that Mrk 573 surely possesses a hidden BLR
in its nucleus. If this kind of misclassification occurs frequently, 
the previous comparative studies would become rather senseless.
Therefore the reason of the misclassification should be discussed here.

Kay (1994) presented the results of the spectropolarimetric observations
for Mrk 573 which were performed in November 1987 -- December 1989.
Although it was mentioned that the H$\beta$ and the H$\gamma$ emission in 
the polarized flux spectrum of Mrk 573 might be slightly broader
than those in the total flux spectrum,
it could not be concluded because the observation did not cover the
H$\alpha$ wavelength range.
It is therefore crucially important for exploring the
hidden BLRs to investigate the profiles of the polarized H$\alpha$ emission,
which is the easiest spectral feature to access the hidden BLR.
It should be mentioned that, however, there may be a possibility that
the scattered BLR emission is temporary variable.
Although there are some reports that the scattered BLR emission of 
type-1 AGNs varies in a few years (e.g., Young et al. 1999;
Smith et al. 2002; Nagao et al. 2004), 
there is no report of finding a significant temporal 
variation of the scattered broad BLR emission in Sy2s, so far.
This is thought to be partly because the scattered photons cannot be 
observed due to the obscuration by a dusty tori if the scattering region
resides at very close to the nucleus, as for Sy2s.
Thus we can see only the polarized light scattered at far from the nucleus
with a distance larger than the scale height of dusty tori, 
$\sim$1 pc or more (e.g., Taniguchi \& Murayama 1998).
Therefore, we cannot detect the temporal variation of scattered BLR emission
of Sy2s unless we monitor the spectropolarimetric data for 
several years at least. As for Mrk 573, the comparison of the
spectropolarimetric data obtained at epochs separated $\sim$15 years may
enable us to access the temporal properties of the scattered BLR emission.
If this is the case, most of the polarization of the hidden BLR in
Mrk 573 is caused at the region within several pc from the nucleus 
where the obscuration by the dusty torus is not significant.

\begin{figure*}
\epsscale{1.18}
\plotone{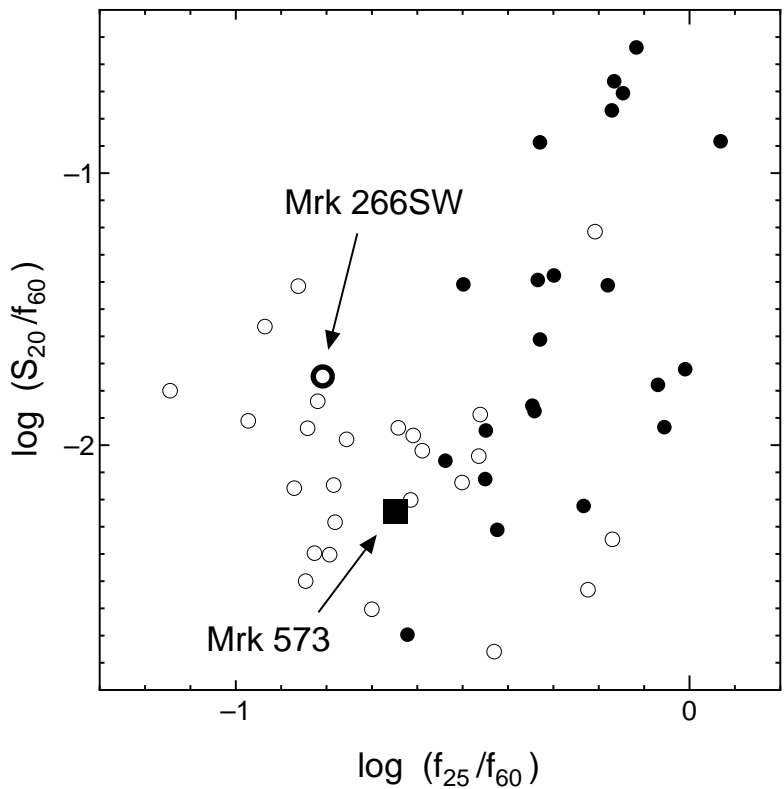}
\caption{
Diagnostic diagram of the ratio of the radio 20cm flux density to
the $IRAS$ 60$\mu$m flux
vs. the $IRAS f_{25}/f_{60}$ color. The data are taken from Tran (2003).
Filled and open circles denote the data of Sy2s with and without a hidden
BLR, respectively. The data of Mrk 573 and Mrk 266SW are shown by a filled
square and a thick circle, respectively.
\label{fig5}}
\end{figure*}

Then we discuss how our finding of the hidden BLR in Mrk 573 affects
the previous studies. It is known that Sy2s with a hidden BLR tend to
show hotter MIR colors (i.e., higher $f_{25}/f_{60}$ ratios where
$f_{25}$ and $_{60}$ are $IRAS$ 25$\mu$m and 60$\mu$m fluxes,
respectively) than Sy2s without a hidden BLR (e.g., Heisler et al. 1997; 
Gu et al. 2001; Tran 2001, 2003). 
Although Heisler et al. (1997) interpreted this tendency
as a difference of viewing angles toward a dusty torus, Alexander (2001)
reported that the $f_{25}/f_{60}$ ratio is not a reliable indicator
for the viewing angle (see also Murayama, Mouri, \& Taniguchi 2000);
rather it denotes the relative strength of the AGN activity compared to
the nuclear star-forming activity. 
In order to compare the relative strength of the AGN activity between Sy2s
with and without a hidden BLR, a diagnostic diagram which
consists of the $f_{25}/f_{60}$ ratio and the $S_{20}/f_{60}$ ratio 
(where $S_{20}$ is a radio 20 cm flux density) has been used since
the $S_{20}/f_{60}$ ratio is also expected to be a good indicator for
the relative AGN activity strength (e.g., Tran 2001, 2003).
In Figure 5, we plot the data of Tran (2003) on the diagnostic diagram 
of $f_{25}/f_{60}$ and $S_{20}/f_{60}$. This is basically the same as
Figure 1 of Tran (2001) (see also Figure 4 of Tran 2003), but the data of
Mrk 573 (and Mrk 266SW, which is discussed below) is explicitly exhibited.
We can see that the data of Mrk 573 is located the edge of the correlation
between the two diagnostic flux ratios for the Sy2s with a hidden BLR.
Actually the $f_{25}/f_{60}$ ratio of Mrk 573 is lower than all of the 
Sy2s with a hidden BLR in the sample of Tran (2003), i.e., 
$f_{25}/f_{60} = 0.23$. Despite the low ratios of $f_{25}/f_{60}$ and 
$S_{20}/f_{60}$, Mrk 573 is thought not to be a peculiar object as a Sy2
with a hidden BLR because its data is consistent with the correlation
between the two diagnostic flux ratios for the Sy2s with a hidden BLR.
The same conclusion can be seen in other diagnostic diagram;
in Figure 6, we show the diagnostic diagram of $f_{25}/f_{60}$ and 
[O {\sc iii}]$\lambda$5007/H$\beta$ in which the data of Tran (2003) are
plotted. As reported by Tran (2001), there is a statistically significant
difference in the distribution of the [O {\sc iii}]$\lambda$5007/H$\beta$
flux ratio between Sy2s with and without a hidden BLR (see also Tran 2003).
Again the data of Mrk 573 is consistent with the data of the Sy2s with
a hidden BLR though it is located at the edge of the distribution of the
data of Sy2s with a hidden BLR in this diagram.
Note that it may be a part of the reason for the undetection of
the scattered BLR emission of Mrk 573 by the previous observations that
the strength of the AGN activity of Mrk 573 is very weak as inferred
by the diagnostic diagrams in Figures 5 and 6.

Our study suggests that high-quality spectropolarimetric observations
are crucial for dividing Sy2s between those with and without a
hidden BLR correctly. In terms of this viewpoint, we
briefly comment on an interesting Sy2, Mrk 266SW.
This Sy2 has been also treated as a Sy2 without a hidden BLR 
(e.g., Tran 2001, 2003) but was suspected to possess the hidden BLR 
by Kay (1994), as well as Mrk 573. Being different from Mrk 573, however,
the diagnostic quantities of Mrk 266SW are not similar to those of the 
Sy2s with a hidden BLR. As seen in Figures 5 and 6, Mrk 266SW shows lower 
[O {\sc iii}]$\lambda$5007/H$\beta$ flux ratio than all of the Sy2s
with a hidden BLR in the sample of Tran (2003), and shows cooler $IRAS$ color
than Mrk 573. The data of Mrk 266SW does not follow the correlation
for the Sy2s with a hidden BLR seen in Figure 5.
Also in the diagnostic diagram of [O {\sc iii}]$\lambda$5007/H$\beta$ vs.
$f_{25}/f_{60}$, the data of Mrk 266SW is far from the data distribution
of the Sy2s with a hidden BLR but is consistent to the Sy2s without a
hidden BLR. Thus we can recognize that the general properties of 
Mrk 266SW is consistent with so-called ``pure Sy2s'', 
i.e., Sy2s without a BLR in its nucleus, not with Sy2s with a BLR.
In order to understand the nature of the Sy2 populations, performing a 
high-quality spectropolarimetry for Mrk 266SW seems important to examine 
whether or not Mrk 266SW possesses a hidden BLR, because Mrk 266SW may be 
an atypical object if possesses a hidden BLR in its nucleus.

\begin{figure*}
\epsscale{1.18}
\plotone{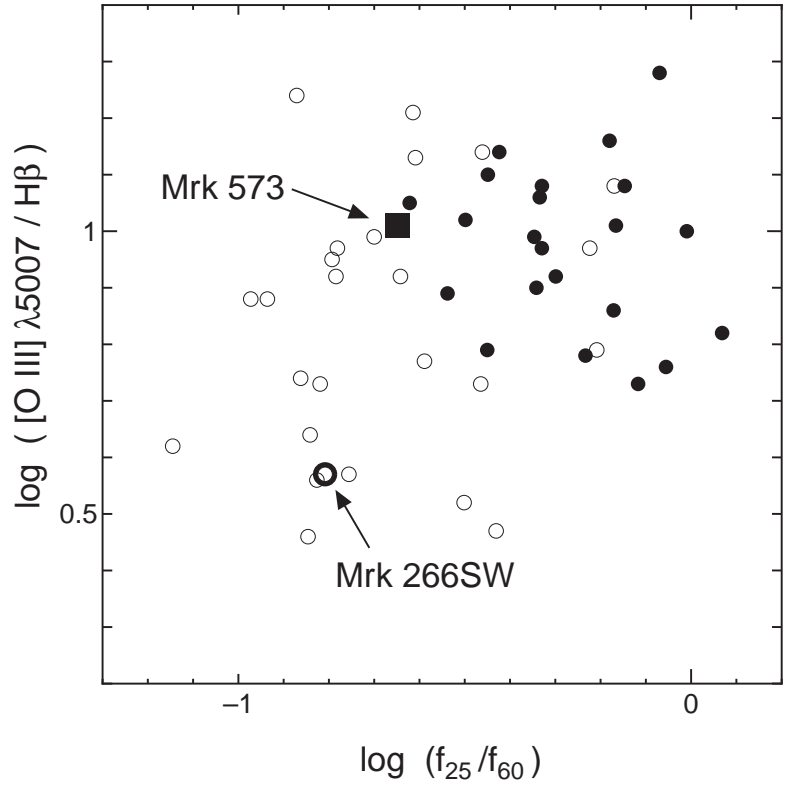}
\caption{
Same as Figure 5 but for other diagnostics; 
the [O {\sc iii}]$\lambda$5007/H$\beta$ flux ratio
vs. the $IRAS f_{25}/f_{60}$ color.
\label{fig6}}
\end{figure*}


\acknowledgments

We are grateful to all the staffs of the 
Subaru telescope, especially to the FOCAS instrument team.
We thank K. Matsuda and M. Seki for useful comments.
TN acknowledges financial support from the Japan Society for the
Promotion of Science (JSPS) through JSPS Research Fellowships for Young
Scientists. A part of this work was financially supported by Grants-in-Aid
for the Scientific Research (10044052, 10304013, and 13740122) of the 
Japanese Ministry of Education, Culture, Sports, Science, and Technology.


\clearpage
\begin{deluxetable}{lcccc}
\tablenum{2}
\tablecaption{Results of the Spectral Fitting around the H$\alpha$ Emission}
\tablehead{
\colhead{Line} &
\colhead{Wavelength\tablenotemark{a}} &
\colhead{Line Width\tablenotemark{b}} &
\colhead{Velocity Width\tablenotemark{c}} &
\colhead{Line Flux\tablenotemark{d}}
}
\startdata  
\cutinhead{Unpolarized Emission Lines}
H$\alpha_{\rm narrow}$    & 
   6563.7 &  9.4 & 307 & 87.20 $\pm$ 0.99 \nl
[N {\sc ii}]$\lambda$6548\tablenotemark{e} & 
   6548.8 & 10.6 & 383 & 23.44 $\pm$ 0.37 \nl
[N {\sc ii}]$\lambda$6583\tablenotemark{e} & 
   6584.1 & 10.7 & 383 & 69.14 $\pm$ 1.09 \nl
\cutinhead{Polarized Emission Lines}
H$\alpha_{\rm broad}$     & 
   6566.0 & 69.8 &         3170 & 2.14 $\pm$ 0.05 \nl
H$\alpha_{\rm narrow}$    & 
   6563.0 &  6.7 & $\lesssim$300\tablenotemark{f} & 0.78 $\pm$ 0.01 \nl
[N {\sc ii}]$\lambda$6548\tablenotemark{e} & 
   6548.3 &  6.0 & $\lesssim$300\tablenotemark{f} & 0.18 $\pm$ 0.00 \nl
[N {\sc ii}]$\lambda$6583\tablenotemark{e} & 
   6583.6 &  6.1 & $\lesssim$300\tablenotemark{f} & 0.53 $\pm$ 0.01
\enddata 
\tablenotetext{a}{Central wavelengths of the Gaussian component in the
                  rest frame of Mrk 573, in units of angstrom.}
\tablenotetext{b}{Measured emission-line widths (FWHM) without the 
                  correction for the instrumental broadening effect, 
                  in units of angstrom.}
\tablenotetext{c}{Velocity widths of the emission lines in units of 
                  km s$^{-1}$. The instrumental broadening is corrected by
                  assuming that the instrumental widths is 300 km s$^{-1}$.}
\tablenotetext{d}{Emission-line fluxes obtained by the best-fit models,
                  in units of 10$^{-15}$ ergs s$^{-1}$ cm$^{-2}$. 
                  The extinction effect is not corrected.}
\tablenotetext{e}{The separation and the flux ratio of the [N {\sc ii}]
                  doublet emission are fixed to be the theoretical values
                  in the fitting process. 
                  The velocity width is fixed to be the same.}
\tablenotetext{e}{Unresolved in the obtained spectrum.}
\end{deluxetable}

\end{document}